\documentclass[12pt]{iopart}
\usepackage{bm}
\usepackage{graphicx}
\usepackage{iopams}
\begin{document}
\title{Diffuse versus square-well confining potentials in modelling $A$@C$_{60}$ atoms}
\author{ V K Dolmatov, J L King and J C Oglesby}
\address{Department of Physics and Earth Science, University of North Alabama,
Florence, Alabama 35632, USA}
\ead{vkdolmatov@una.edu}
\begin{abstract}
Attention: this version-$2$ of the manuscript differs from its previously uploaded version-$1$ (arXiv:1112.6158v1) and subsequently published in 2012 J. Phys. B \textbf{45} 105102 only by a removed typo in Eq.(2) of version-$1$; there was the erroneous factor ``2'' in both terms in the right-hand-side of the Eq.(2) of version-$1$. Now that the typo is removed, Eq.(2) is correct.

A perceived advantage for the replacement of a discontinuous square-well pseudo-potential, which is often used by various researchers as an approximation to the actual C$_{60}$ cage potential in calculations of endohedral atoms $A$@C$_{60}$, by a more realistic diffuse potential is explored. The photoionization of endohedral H@C$_{60}$ and Xe@C$_{60}$ is chosen as the case study. The diffuse potential is modelled by a combination of two Woods-Saxon potentials. It is demonstrated that photoionization spectra of $A$@C$_{60}$ atoms are largely insensitive to the degree $\eta$ of diffuseness of the potential borders, in a reasonably broad range of $\eta$'s. Alternatively, these spectra are found to be insensitive to discontinuity of the square-well potential either. Both potentials result in practically identical calculated spectra. New numerical values for the set of square-well parameters, which lead to a better agreement between experimental and theoretical data for $A$@C$_{60}$ spectra, are recommended for future studies.
\end{abstract}
%\submitto{\jpb}
\maketitle
\section{Introduction}
\label{intro}
Photoionization of atoms $A$ encapsulated inside the hollow interior of the C$_{60}$ fullerene cage,
labelled as $A$@C$_{60}$ and referred to as endohedral or confined atoms, has been an \textit{ad hoc} topic
of numerous theoretical (see review papers \cite{DolmatovAQC,DolmRPC04} (and references therein
in addition to other references in the present paper) and experimental \cite{Kilcoyne} (and references therein) studies in recent years.
An \textit{ab initio}, based on first principles theory of $A$@C$_{60}$ photoionization
has not been developed yet, in view of a formidable complexity of the problem. Perhaps, the most sophisticated theory here is
represented by a time-dependent local density approximation (TDLDA), see, e.g., \cite{Himadri} and references therein. However, the current TDLDA
theory has obvious drawbacks in view of an unsatisfactory agreement with the existing experimental data on the Xe@C$_{60}^{+}$ $4\rm d$ photoionization
\cite{Kilcoyne}. Meanwhile, many important insights into the problem can be, and have been, unravelled on the basis
of simpler empirical models based on the modelling of C$_{60}$ confinement by pseudo-potentials, such as a $\delta$-function-like potential
\cite{AmusiaXe@C60,Galina} (and references therein) or square-well potential, $U_{\rm SWP}(r)$. The latter is defined as
\begin{eqnarray}
 U_{\rm SWP}(r)=\left\{\matrix {
- U_{0}, & \mbox{if $R_{0} \le r \le R_{0}+\Delta$} \nonumber \\
0 & \mbox{otherwise.} } \right.
\label{SWP}
\end{eqnarray}
Here, $R_{0}$ is the inner radius of C$_{60}$, $\Delta$ is the thickness of the C$_{60}$ wall, and $U_{0}$ is the potential depth.
The square-well potential modelling of C$_{60}$ has become quite popular among various researchers. It has been used on numerous occasions
in the field of C$_{60}$ and $A$@C$_{60}$ theoretical studies, see, e.g.,
\cite{DolmatovAQC,DolmRPC04,Pushka,Xu,Roth,Solovyov,Iran,Ndengu,CCR,MitnikPRA08,Pranawa,Ludlow,Mitnik,Grum,Chen} and references therein.

The potential $U_{\rm SWP}(r)$, however, is discontinuous at its borders. A possible emergence of qualitative and, especially, quantitative artifacts in calculated $A$@C$_{60}$ spectra, due to said discontinuity,
in calculated photoionization cross sections has not been detailed in literature except for a couple of brief remarks dropped on the subject in passing \cite{Fourier1,Wendin}.
In particular, it is because of
the lack of such knowledge that it has recently been proposed \cite{Prudente} to replace the square-well potential modelling of the C$_{60}$  by a smooth
Gaussian-function-like model potential. However, the latter continuously changes strongly everywhere
inside the C$_{60}$ cage, thereby having no compact borders. This contradicts a recent Fourier imaging study of the experimental C$_{60}$ photoionization cross section \cite{Fourier2}. The study shows
that C$_{60}$ has well-defined, compact borders. In short, following
the logic line of \cite{Fourier1,Fourier2}, photoionization of an atomic cluster occurs with the greatest probability where the cluster's potential
changes sharply. This is obvious from the acceleration form gage for a dipole photoionization amplitude, $D$, namely, $D \propto (\psi_{f}|\Delta_{r}V(r)|\psi_{i})$. Based on this, it was determined \cite{Fourier2}
 that C$_{60}$ has sharp borders rather than soft borders. Correspondingly, the assumption for a Gaussian-function-like potential of C$_{60}$ is incorrect.

Surely, a C$_{60}$ confining potential with diffuse (rounded) but compact borders, if defined appropriately,  is more realistic than a square-well potential with infinitely sharp edges. However,
it is not at all clear beforehand to what degree replacement of the square-well potential for a diffuse potential may improve or worsen agreement between experiment and theory.
The need for clarifying this  issue is signified by the fact that the square-well potential modelling of $A$@C$_{60}$ atoms has been used abundantly over the years. This has resulted
in a large array of predicted data and phenomena that might need to be re-studied/re-calculated with an eye on more realistic diffuse potential borders.
It is the aim of this paper to dot \textit{i}'s and cross \textit{t}'s regarding the stated concerns.

To attain the aim, we calculate the photoionization cross sections of, and photoelectron angular distribution from, inner and valence subshells of $A$@C$_{60}$ atoms by utilizing both $U_{\rm SWP}(r)$  and
$U_{\rm DP}(r)$ as the C$_{60}$ confining potentials. As a result, we find that discontinuity of the  $U_{\rm SWP}(r)$ potential leads neither to qualitative nor quantitative artifacts in calculated photoionization characteristics of $A$@C$_{60}$ atoms compared to calculated data obtained with the use of the
diffuse potential $U_{\rm DP}(r)$. Moreover, it is found that the degree of diffuseness of
the $U_{\rm DP}(r)$ potential matters surprisingly little - a fraction-to-nothing - in a relatively broad range of its values.

Atomic units (\textit{au}) are used throughout this paper.
\section{Theory, Results, and Discussion}
\label{sec:1}
\subsection{Review of theory}
\label{sec:1.1}

In this work, a square-well potential $U_{\rm SWP}(r)$ is given by (\ref{SWP}). As for a diffuse potential $U_{\rm DP}(r)$, it is defined by a combination of two Woods-Saxon potentials:
\begin{eqnarray}
\fl U_{\rm DP}(r) & = \left. \frac{U_{0}}{1+{\exp}(\frac{R_{0} -r}{\eta})}\right |_{r \le R_{0} +\frac{1}{2}\Delta} \nonumber\\
\fl & +\left. \frac{U_{0}}{1+{\exp}(\frac{r-R_{0}-\Delta}{\eta})}\right |_{r > R_{0}+\frac{1}{2}\Delta}.
\label{DP}
\end{eqnarray}
Here, $\eta$ is the diffuseness parameter, and $R_{0}$, $U_{0}$, and $\Delta$ are the same as the parameters of the square-well potential.

Concerning photoionization of $A$@C$_{60}$ atoms, we focus on the photoionization cross section $\sigma_{n \ell}(\omega)$ of a $n\ell$-subshell  of $A$@C$_{60}$ as well as dipole photoelectron angular-asymmetry
parameter $\beta_{n\ell}(\omega)$. For free atoms, they are determined by equations presented, e.g., in \cite{AmusiaATOM,Cooper93}. The latter are equally applicable to endohedral atoms  $A$@C$_{60}$ as well, in
our modelling of such atoms. Correspondingly,
\begin{eqnarray}
 \sigma_{n \ell} = \frac{4\pi{^2}\alpha N_{n \ell}}{3(2\ell+1)} \omega [\ell
|D_{\ell-1}|^{2}+ (\ell+1)|D_{\ell+1}|^{2}],
\label{eqCC}
\end{eqnarray}
and we prefer to recast $\beta_{n\ell}(\omega)$ as
\begin{eqnarray}
 \beta_{n \ell} =
\frac
{
\ell (\ell-1)\rho^{2} - 6\ell(\ell+1)\rho\cos\Phi
+ (\ell +1)(\ell+2)}
{(2\ell+1)(\rho^{2}\ell +\ell +1)
}.
\label{eqbeta2}
\end{eqnarray}
Here,
\begin{eqnarray}
\rho = \frac{|D_{\ell-1}|}{|D_{\ell+1}|}, \quad \Phi=\delta_{\ell+1}-\delta_{\ell-1}.
\label{eqrho}
\end{eqnarray}
In the above equations, $\omega$ is the photon energy, $N_{n\ell}$ is the
number of electrons in a $n\ell$ subshell, $\alpha$ is the fine-structure constant,
$D_{\ell\pm 1}$ are radial dipole photoionization amplitudes, and $\delta_{\ell \pm 1}$ are phase shifts of $D_{\ell\pm 1}$.
 Note that the quantities $\sigma_{n\ell}$, $\beta_{n\ell}$, $D_{\ell\pm 1}$,
  $\rho$, $\delta_{l\pm1}$, and
$\Phi$ all depend upon photon energy $\omega$; the explicit dependence
is omitted in the above equations for reasons of simplicity.

In an independent particle approximation, $D_{\ell\pm 1}$ is defined as
\begin{eqnarray}
D_{\ell\pm 1}=\int_{0}^{\infty}{P_{\epsilon \ell\pm 1}(r) r P_{n\ell}(r) dr},
\label{eqD}
\end{eqnarray}
where $P_{n \ell}(r)$ and $P_{\epsilon \ell}(r)$ are one-electron radial wavefunctions of the bound and continuous states, respectively.
In the present work, these wavefunctions and energies $E_{n\ell}$ of a discrete spectrum are the solutions of a modified radial Hartree-Fock (HF) equation accounting for a $U_{\rm C_{60}}(r)$ confinement:
\begin{eqnarray}
[\hat{H}^{r} + U_{\rm C_{60}}(r)]P_{n\ell}(r) = E_{n(\epsilon)\ell}P_{n(\epsilon)\ell}(r).
\label{HF}
\end{eqnarray}
Here, $\hat{H}^{r}$ is a radial part of the HF Hamiltonian which is identical to that for a free atom \cite{AmusiaATOM}, and
$U_{\rm C_{60}}(r)$ is either the square-well potential $U_{\rm SWP}(r)$ (\ref{SWP}), or diffuse potential $U_{\rm DP}(r)$ (\ref{DP}).

In a particular case of an endohedral single-electron hydrogen atom, H@C$_{60}$, (\ref{HF}) reduces to a radial
Schr\"{o}dinger equation in the presence of the C$_{60}$ confinement:
\begin{eqnarray}
\fl -\frac{1}{2}\frac{d^2P_{n(\epsilon) \ell}}{dr^2} +\left [\frac{-1}{r} +\frac{\ell(\ell+1)}{2 r^2} +U_{\rm C_{60}}(r)\right ]P_{n(\epsilon)\ell}(r) \nonumber \\
\fl = E_{n(\epsilon) \ell} P_{n(\epsilon)\ell}(r).
\label{EqH}
\end{eqnarray}

To calculate $D_{\ell\pm 1}$ beyond the independent particle HF approximation, i.e., to account for electron correlation,
we utilize a random phase approximation with exchange (RPAE) \cite{AmusiaATOM}.

\subsection{Results and discussion}
\label{sec:1.2}
In this work, when approximating the C$_{60}$ potential by a square-well or diffuse potential, we
use $U_{0}= -0.422$, $\Delta = 1.25$, and $R_{0}=6.01$ \textit{au} \cite{Keating} rather than
$U_{0}= -0.302$, $\Delta = 1.9$, and $R_{0}=5.89$ \textit{au} used in previous calculations by these and many other authors, see, e.g., \cite{DolmatovAQC,DolmRPC04,CCR}.
This is because the new parameters were shown \cite{Keating} to result in a
much better match of calculated photoionization spectra to
experimental spectra of endohedral atoms, particularly the $\rm 4d$ spectrum of Xe@C$_{60}^{+}$ \cite{Kilcoyne}.

\begin{figure}[h]
\center{\includegraphics[width=8cm]{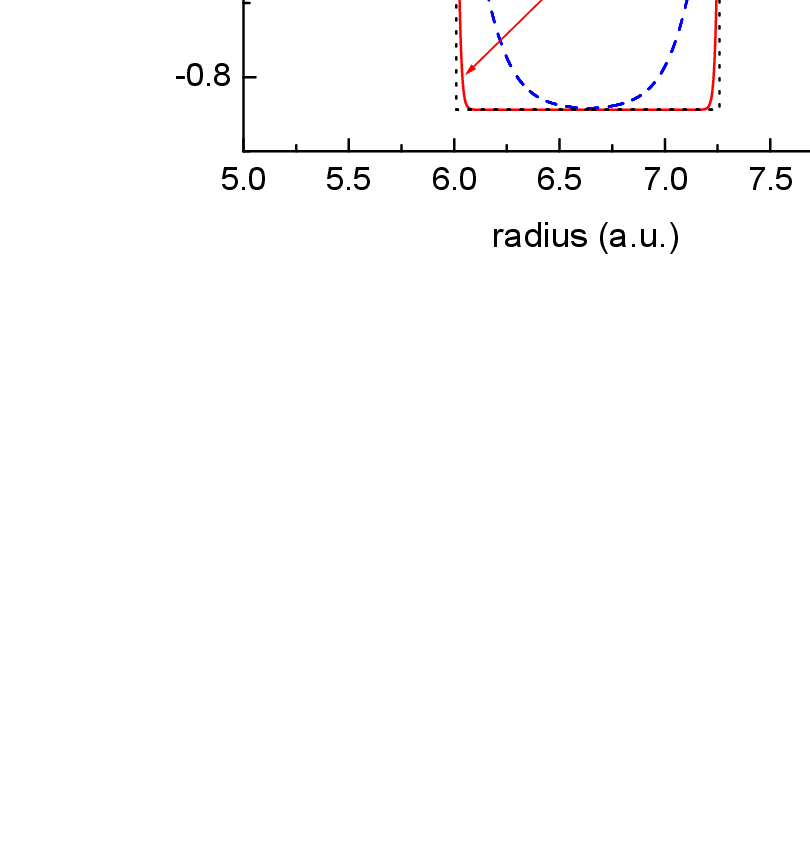}}
\caption{The  square-well potential $U_{\rm SWP}(r)$ (\ref{SWP}) with  $U_{0}= -0.422$, $\Delta = 1.25$, and $R_{0}=6.01$ \textit{au} (dotted line),
as well as the diffuse  potential $U_{\rm DP}(r)$ (\ref{DP}) with the diffuseness parameter $\eta=0.01$ (solid line) and $\eta=0.1$ (dashed line). See text for details.}
\label{fig1}
\end{figure}

Calculated data for thus defined $U_{\rm SWP}(r)$ and $U_{\rm DP}(r)$ are compared with each other in figure~\ref{fig1} for two different values of the
diffuseness parameter $\eta$ of $U_{\rm DP}(r)$, namely, $\eta=0.01$ and $0.1$ .
One can see from figure~\ref{fig1} that $U_{\rm DP}(r)$ with $\eta=0.01$ and the square-well potential are practically identical. On the contrary, for $\eta=0.1$, calculated $U_{\rm DP}(r)$
has strongly diffuse borders, thereby noticeably deviating from the square-well potential.

In the following, we detail results of comparison between  $\sigma_{n \ell}(\omega)$ and $\beta_{n\ell}(\omega)$ calculated with the use of the $U_{\rm SWP}(r)$  and $U_{\rm DP}(r)$
confining potentials.

\begin{figure}
\center{\includegraphics[width=8cm]{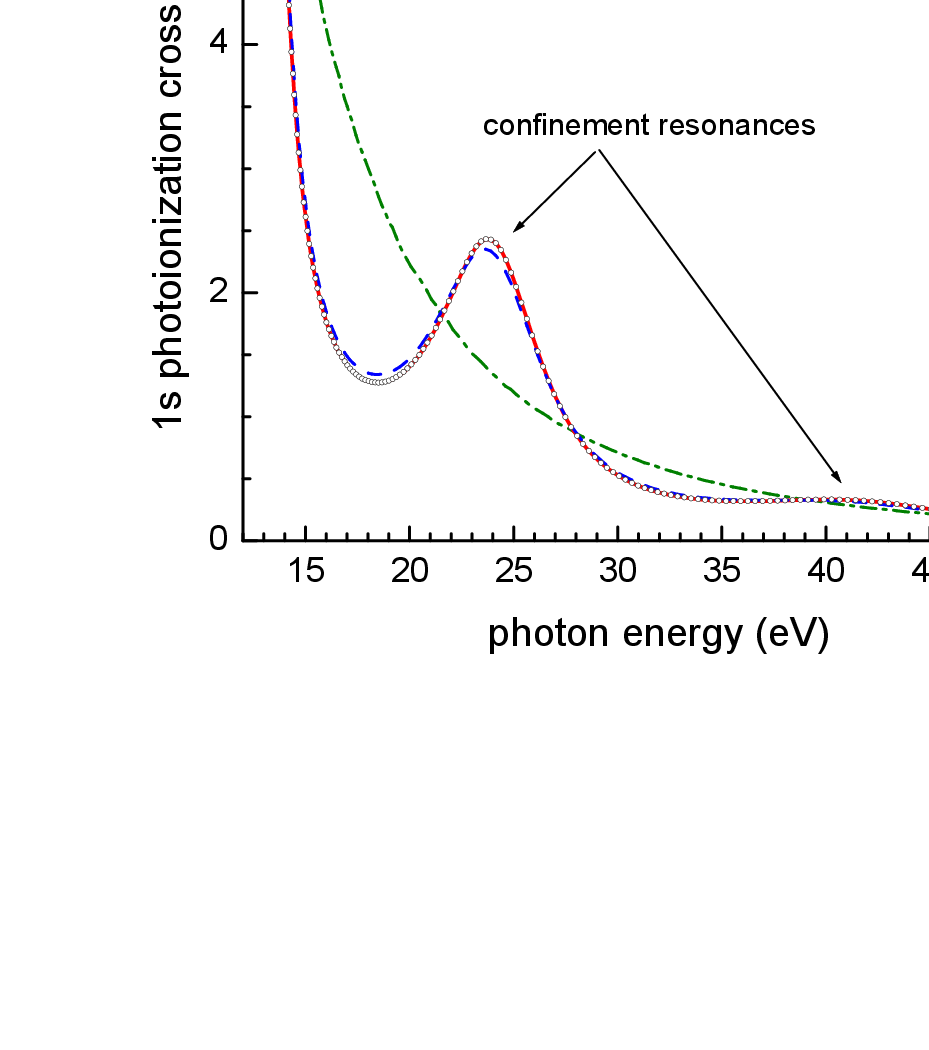}}
\caption{Hydrogen $1\rm s$ photoionization cross section of H@C$_{60}$ calculated with the use of the square-well potential $U_{\rm SWP}(r)$ (open circles),
 diffuse potential $U_{\rm DP}(r)$ with $\eta=0.01$ (solid line) and $\eta=0.1$ (dashed line), and of free hydrogen, as marked.}
\label{fig2}
\end{figure}

As the first step, in order to avoid various possible complications
due to electron correlation in multielectron atoms,  we discuss a ``clean'' case - the photoionization of the $1\rm s$ ground-state of the endohedral hydrogen atom, H@C$_{60}$.
Corresponding results are depicted in figure~\ref{fig2}.

First, we note that the photoionization cross section $\sigma_{\rm 1s}(\omega)$ of H@C$_{60}$, calculated with the use of a diffuse potential $U_{\rm DP}(r)$ with
a small diffuseness parameter $\eta=0.01$, is practically undistinguishable from $\sigma_{\rm 1s}(\omega)$ calculated in the square-well potential modelling of H@C$_{60}$.
Hence, discontinuity of the square-well potential does not result in any artifacts in $\sigma_{\rm 1s}(\omega)$. Second, what comes as a big surprise, is
that calculated $\sigma_{\rm 1s}(\omega)$, obtained with the use of a much greater diffuseness parameter $\eta=0.1$, is practically
identical to $\sigma_{\rm 1s}(\omega)$ calculated either with $\eta=0.01$ or in the framework of the square-well potential model. This speaks to the fact that the photoionization
 of an endohedral atom may be rather insensitive to the degree of diffuseness of the confining potential, in reasonably large limits. As a note on an independent issue, one can see that the photoionization cross section of free hydrogen differs from that
 of H@C$_{60}$ by the presence of a well developed maximum at about $25$ eV and insignificant maximum at about $41$ eV in the cross section. These are called confinement resonances, i.e., resonances that occur due to a constructive interference
 of the outgoing photoelectron wave and photoelectron waves scattered off the C$_{60}$ cage \cite{DolmatovAQC,DolmRPC04,AmusiaXe@C60} (and references therein).
  Confinement resonances have only recently been experimentally proven to exist \cite{Kilcoyne}.

\begin{figure}[h]
\center{\includegraphics[width=8cm]{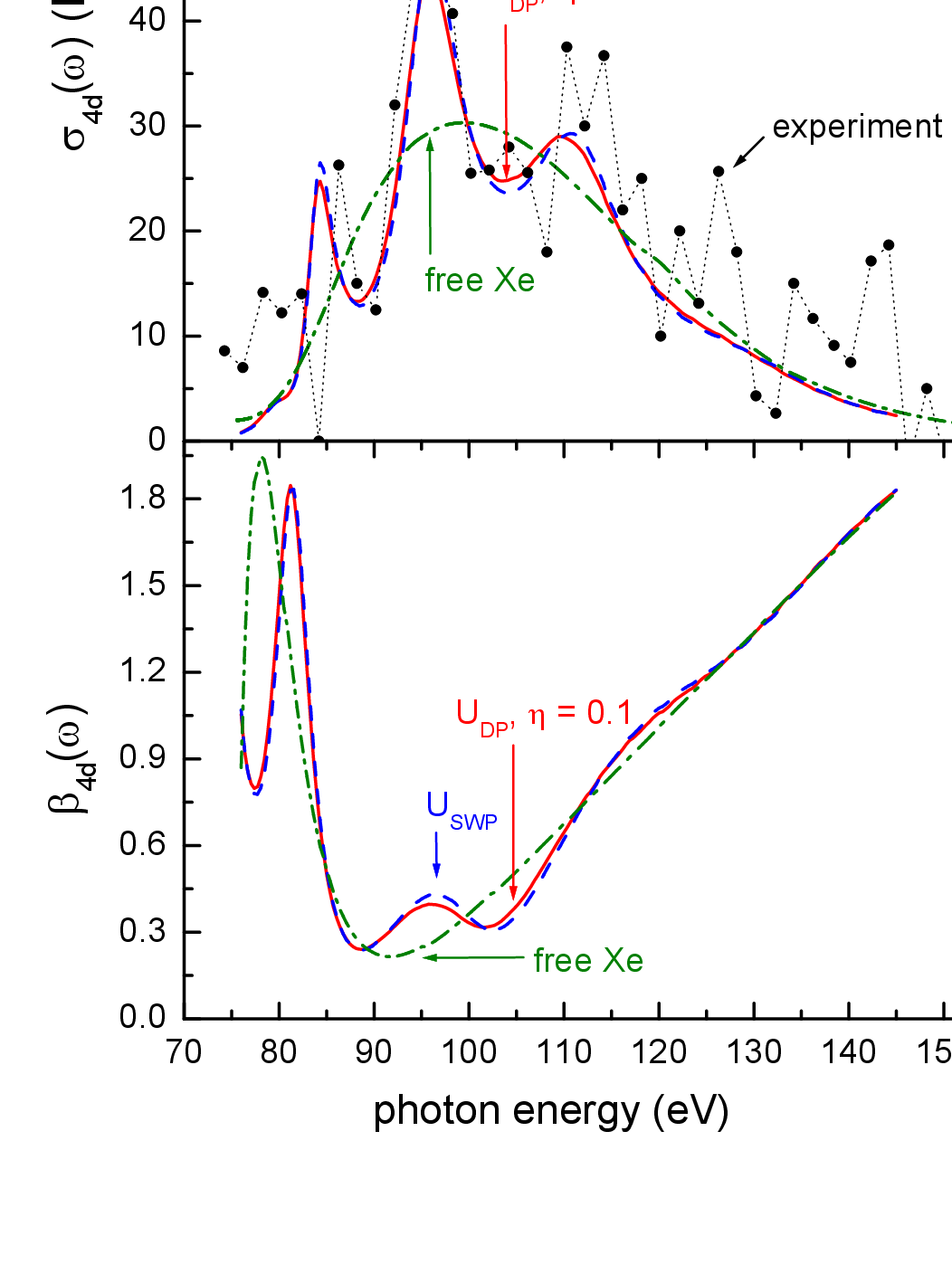}}
\caption{RPAE calculated data for $\sigma_{\rm 4d}(\omega)$ and $\beta_{\rm 4d}(\omega)$ of Xe@C$_{60}$
obtained with the use of the square-well potential $U_{\rm SWP}(r)$ (dashed lines) and diffuse potential $U_{\rm DP}(r)$ with $\eta=0.1$ (solid lines), as marked.
 Also depicted are corresponding RPAE calculated data for free Xe (dashed-dotted lines) and
 experimental data for $\sigma_{\rm 4d}(\omega)$ of Xe@C$_{60}^{+}$ \cite{Kilcoyne}. Experimental data were shifted by $4.2$ eV towards higher photon
 energies as well as multiplied by $10$ to compare with theory (see discussion in Appendix).}
\label{fig3}
\end{figure}

\begin{figure}[h]
\center{\includegraphics[width=8cm]{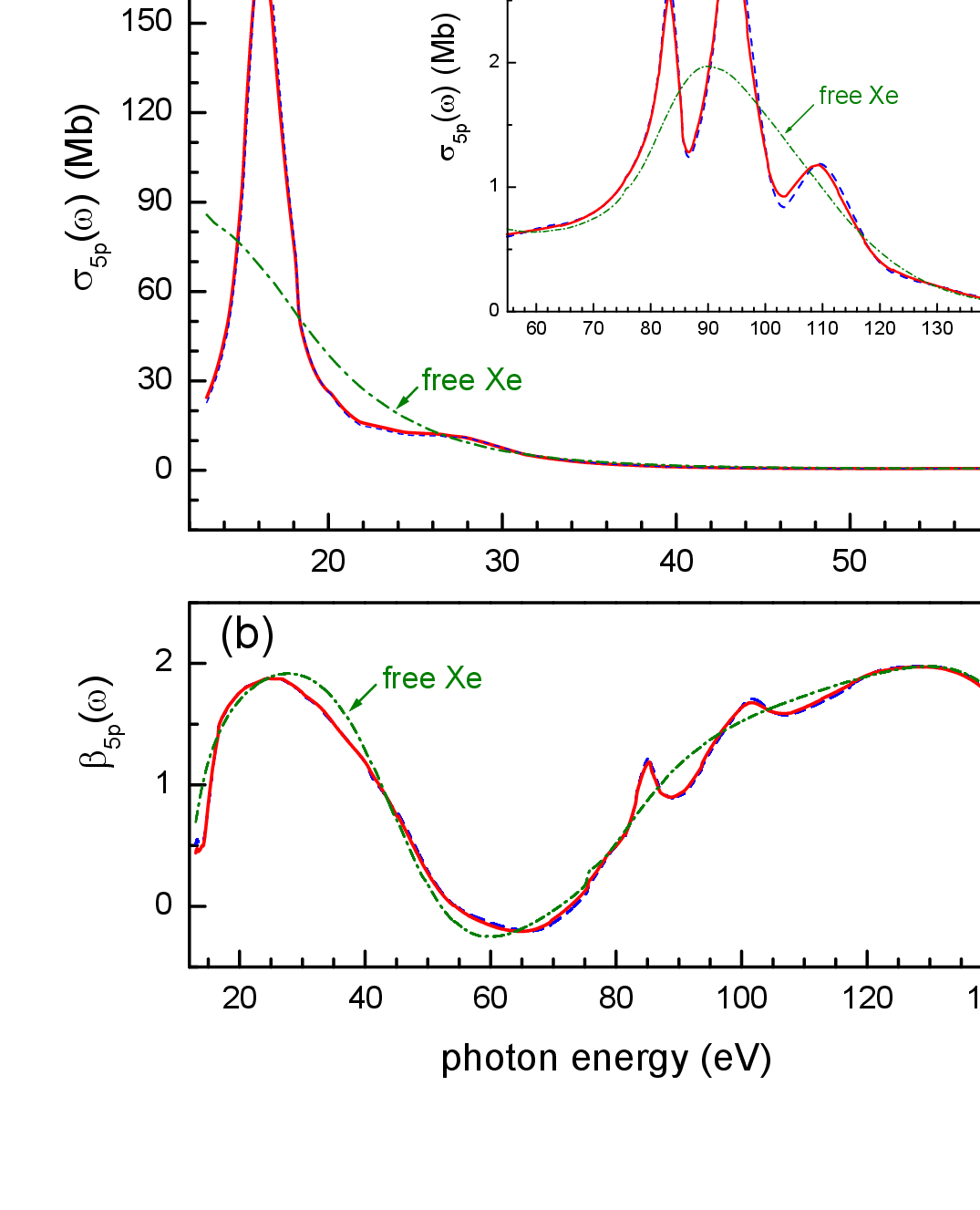}}
\caption{RPAE calculated data for Xe@C$_{60}$ $\rm 5p$ photoionization cross section, $\sigma_{\rm 5p}(\omega)$, and dipole photoelectron angular-asymmetry parameter, $\beta_{\rm 5p}(\omega)$,
calculated with the use of the square-well potential $U_{\rm SWP}(r)$ (dashed lines) and diffuse potential $U_{\rm DP}(r)$ with $\eta=0.1$ (solid lines). Also depicted are RPAE calculated data for
$\sigma_{\rm 5p}(\omega)$ and $\beta_{\rm 5p}(\omega)$ for free Xe (dashed-dotted lines).}
\label{fig4}
\end{figure}

Next, we investigate whether the above findings are valid for photoionization of inner-shells and valence-shells of  endohedral multielectron atoms $A$@C$_{60}$ where
electron correlation is strong. To date, only photoionization of endohedral Xe@C$_{60}^{+}$ has been
 reliably measured experimentally \cite{Kilcoyne}. Therefore, we choose photoionization of Xe@C$_{60}$ as the case study to learn
 how discontinuity of $U_{\rm SWP}(r)$ or diffuseness of $U_{\rm DP}(r)$ may affect photoionization cross sections $\sigma_{n\ell}(\omega)$ of, and photoelectron angular-asymmetry parameters $\beta_{n\ell}(\omega)$ from,  the Xe@C$_{60}$ inner ${\rm 4d}^{10}$ and
 valence ${\rm 5p}^{6}$  subshells. The comparison of our calculated data for neutral Xe@C$_{60}$ with experimental data for charged Xe@C$_{60}^{+}$ \cite{Kilcoyne} is appropriate because \cite{DolmatovAQC,Coulumb} charging the C$_{60}$ shell positively does nothing
to the photoionization cross section (as a function of photon energy) except to increase the threshold energy. Note, $\beta_{n\ell}(\omega)$ depends on phase shifts $\delta_{\ell\pm 1}$ of photoionization matrix elements. Hence, corresponding calculated data
will provide, although implicitly,  the information on sensitivity of phase shifts $\delta_{\ell\pm 1}$ to diffuseness of a confining potential as well.  Both HF and RPAE calculated data for $\sigma_{\rm 4d}(\omega)$
and $\beta_{\rm 4d}(\omega)$, as well as for $\sigma_{\rm 5p}(\omega)$
and $\beta_{\rm 5p}(\omega)$, are depicted in figures~\ref{fig3} and \ref{fig4}, respectively.

One can see from figures~\ref{fig3} and \ref{fig4} (compare dashed curves marked as ``$U_{\rm SWP}$'' with solid curves marked as ``$U_{\rm DP}$, $\eta=0.1$'') that
calculated data for $\sigma_{n\ell}(\omega)$ and $\beta_{n\ell}(\omega)$, obtained with the use of either of the two potentials, are practically identical even in the presence
of a strong electron correlation in the atom. This leads us to the conclusions outlined in the next section.

\section{Conclusion}
\label{sec:2}

To summarize, it has been proven in this paper that discontinuity
of a square-well potential does not lead to any artifacts, whether quantitative or qualitative, in photoionization spectra of endohedral atoms. The square-well potential modelling of C$_{60}$ is as good as
its modelling by a diffuse potential. Both of these potentials lead to practically identical calculated data for photoionization spectra. The latter are largely insensitive to the degree of diffuseness.
Hence, both potentials work equally when approximating the C$_{60}$ cage potential. The implication is that replacing the square-well potential by  a more realistic diffuse potential,
if wanted, will not require a re-study of a rich variety of important results on $A$@C$_{60}$ ionization phenomena accumulated to date on the basis of the square-well confining potential concept.

Finally, since the used here values of the square-well potential parameters $R_{0}=6.01$, $\Delta = 1.25$ and $U_{0}=0.422$ \textit{au} result in a reasonable agreement
with experiment (see figure \ref{fig3}), we suggest the users of the square-well potential modelling of $A$@C$_{60}$ to utilize these updated parameters in their future work on the subject.

\ack
The authors are thankful to Professor M. Ya. Amusia for helpful discussions. This work was supported by NSF Grant No.\ PHY-$0969386$.

\section*{Appendix: The $\rm 4d$ and $\rm 5p$ photoionization of Xe@C$_{60}$}

To finalize the paper, we now briefly comment on particular features and structures in the calculated $\rm 4d$ and $\rm 5p$ spectra of Xe@C$_{60}$ depicted in figures~\ref{fig3} and \ref{fig4}, since they are important stand-alone qualities.

To start with, the three strong oscillations in
$\sigma_{\rm 4d}(\omega)$ of Xe@C$_{60}$ about  $\sigma_{\rm 4d}(\omega)$ of free Xe are confinement resonances. They were previously theoretically
studied and detailed in \cite{Himadri,AmusiaXe@C60,Pranawa,CCR,Keating} (and references therein). Experimentally, confinement resonances in $A$@C$_{60}$ photoionization were observed only recently, in the $4d$ giant resonance spectrum of Xe@C$_{60}^{+}$, as the case study \cite{Kilcoyne}. Results of the measurement are by about a factor of $10$ smaller than the cited theoretical
data.  This is because  the measured channel of the Xe@C$_{60}^{+}$ $\rm 4d$ photoionization might  account for only about $10\%$ of the Xe $\rm 4d$ oscillator strength \cite{Kilcoyne}. For this reason, to ease the comparison
with the presented herein theoretical data, the experimental data were multiplied by $10$, and so depicted in figure~\ref{fig3}. One can see that calculated results for $\sigma_{\rm 4d}(\omega)$ of Xe@C$_{60}$, obtained with the
use of $R_{0}=6.01$, $\Delta = 1.25$ and $U_{0}=0.422$ \textit{au}, are in a reasonable agreement with main structures of the experimental spectrum, to a good approximation. Of course it would be too naive to expect our simple model to account for all qualitative and quantitative features of the measured Xe@C$_{60}^{+}$ photoionization which (the model), among other things, accounts for only a single-electron $4d$ ionization. Thus, for example,
while confinement brought oscillations in the experimental spectrum definitely emerge from the single-electron photoionization, possible double-electron satellites might affect their positions and amplitudes as well. We, however, assume that such contributions are not decisive, so that omitting them, as in the present work, is a reasonable first step towards the initial understanding of the experimental data.

To conclude the discussion on the Xe@C$_{60}$ $\rm 4d$ photoionization we note that, recently, another group of theorists \cite{Chen} has reported results of their own calculations of the Xe@C$_{60}$ $\rm 4d$ photoionization. They used exactly the same approximations as three other \textit{independent} theoretical groups \cite{Pranawa,CCR,Private} [i.e., RPAE (or relativistic RPAE \cite{Pranawa}) \& square-well confining
potential with identical parameters). However, results of work \cite{Chen} differ strongly from results of all the three cited theoretical groups (the latter are in a good agreement with each other). We conclude that
 something was not right in work \cite{Chen}, most likely because of a very peculiar way those authors chose for solving HF equations in the presence of a square-well potential confinement.

 Commenting on $\sigma_{\rm 5p}(\omega)$  of Xe@C$_{60}$, see figure~\ref{fig4}~(a), we note that a strong maximum at the $5p$ threshold as well as a less developed oscillation at about $28$ eV in $\sigma_{\rm 5p}(\omega)$
 are confinement resonances. Interesting, they are absent in $\beta_{\rm 5p}(\omega)$. This is because the $\rm 4d$ $\rightarrow$ $\rm f$ transition absolutely dominates over the $\rm 4d$ $\rightarrow$ $\rm p$ transition at given energies both for free Xe and Xe@C$_{60}$. Correspondingly, the parameter $\rho$ (\ref{eqrho}) is negligible, $\rho << 1$, in both atoms. As a result, it vanishes from the equation for $\beta_{n\ell}$
 [see (\ref{eqbeta2})], and so do confinement resonances in question. At higher energies, above $70$ eV,
 $\sigma_{\rm 5p}(\omega)$ is dominated by three strong resonances. Their positions approximately match the positions of confinement resonances
 in $\sigma_{\rm 4d}(\omega)$, see figure~\ref{fig3}. The same situation has recently  been found in the Xe@C$_{60}$ $\rm 5p$ generalized oscillator strength \cite{Amusia_e-Xe}. As in \cite{Amusia_e-Xe}, the present study reveals that the resonances in $\sigma_{\rm 5p}(\omega)$ beyond $70$ eV are induced by the confinement resonances in the $\rm 4d$ photoionization channels, via interchannel coupling. Thus,
 the resonances in $\sigma_{\rm 5p}(\omega)$ beyond $70$ eV are \textit{correlation} confinement resonances \cite{DolmatovAQC,Himadri,CCR,Pranawa,ResurrectedCRs}.  Correlation confinement resonances in $A$@C$_{60}$ photoionization
 are  resonances that emerge in the photoionization of an outer subshell due to interference of transitions
from the outer subshell ($\rm 5p$ in our case) with confinement resonances emerging in \textit{inner} shell
transitions ($\rm 4d$ transitions in our case), via interchannel coupling.
Next, one can see from figure~\ref{fig4}~(b) that these correlation confinement resonances
emerge in $\beta_{\rm 5p}(\omega)$ as well, at about $85$ and $100$ eV. This is because the otherwise dominant $\rm 4d$ $\rightarrow$ $\rm f$ transition is minimized at these energies, as clearly follows from data
for $\sigma_{\rm 5p}(\omega)$. The amplitudes of $\rm 4d$ $\rightarrow$ $\rm f$ and $\rm 4d$ $\rightarrow$ $\rm p$ transitions become comparable, so that $\rho \sim 1$, in contrast to $\rho << 1$ when
$\rm 4d$ $\rightarrow$ $\rm f$ is maximized. Correspondingly, the parameter $\rho$ noticeably oscillates through the photon energy region of $70$ to $110$ eV, and so does $\beta_{\rm 5p}(\omega)$. This is
why noticeable confinement resonances emerge in $\beta_{\rm 5p}(\omega)$ above $70$ eV, in contrast to their absence near threshold.
Finally, we note that we ignored the C$_{60}$ plasmon resonance impact \cite{Himadri,Solovyov,GiantCR} (and references therein) on the Xe@C$_{60}$ $\rm 5p$ ionization, although it is known to be significant
up to about $70$ eV of photon energy. The omission  is justified because the
primary aim of the present paper is to study relative differences between the effects of the  diffuse and square-well confining potentials on $A$@C$_{60}$ photoionization. The C$_{60}$ plasmon resonances cannot alter
these differences. For the same reason, we omitted accounting for the existing, but generally weak, interior static-polarization effect in $A$@C$_{60}$ \cite{StaticE}.
\section*{References}

\end{document}